\documentclass[flushrt]{aastex}
\usepackage{emulateapj5}
\received{2000 October 13}
\revised{2001 April 3}
\accepted{2001 April 17}
\journalinfo{Astrophysical Journal, in press.  astro-ph/0105035}
\submitted{Received 2000 October 13; accepted 2001 April 17.}

\usepackage{ifthen}
\usepackage{natbib}

\shorttitle{SL9 Impact I: Plume and Images}
\shortauthors{Harrington and Deming}

\newcommand\vmax{v\sb{\mbox{\scriptsize max}}}
\newcommand\vmin{v\sb{\mbox{\scriptsize min}}}
\newcommand\mvmax{\math{v\sb{\mbox{\scriptsize max}}}}
\newcommand\mvmin{\math{v\sb{\mbox{\scriptsize min}}}}
\newcommand\mvan{m\sb{\mbox{\scriptsize van}}}
\newcommand\mplume{m\sb{\mbox{\scriptsize plume}}}
\newcommand\mdiff{{\mathrm{d}}}

\newcommand\degree{\degr}
\newcommand\degrees\degree
\font\greek = psyr
\newcommand\micro{{\greek m}}
\renewcommand\micron{\micro m}
\newcommand\microns \micron
\let\oldsim=\sim
\renewcommand\sim{\ifmmode\oldsim\else\math{\oldsim}\fi}
\let\oldpm=\pm
\renewcommand\pm{\ifmmode\oldpm\else\math{\oldpm}\fi}
\newcommand\by{\ifmmode\times\else\math{\times}\fi}
\newcommand\ttt[1]{10\sp{#1}}
\newcommand\tttt[1]{\by\ttt{#1}}
\newcommand\tablebox[1]{\begin{tabular}[t]{@{}l@{}}#1\end{tabular}}

\newcommand\citeauthyear[1]{\citeauthor{#1} \citeyear{#1}}

\catcode`@=11

\renewcommand\comment[1]{}

\newcommand\commenton{\catcode`\%=14}
\newcommand\commentoff{\catcode`\%=12}

\renewcommand\math[1]{$#1$}
\newcommand\mathshifton{\catcode`\$=3}
\newcommand\mathshiftoff{\catcode`\$=12}

\comment{the backslash is necessary}

\comment{alignment tab}

\let\atab=&
\newcommand\atabon{\catcode`\&=4}
\newcommand\ataboff{\catcode`\&=12}

\let\oldmsp=\sp
\let\oldmsb=\sb
\renewcommand\sp[1]{\ifmmode
	   \oldmsp{#1}%
	 \else\strut\raise.85ex\hbox{\scriptsize #1}\fi}
\renewcommand\sb[1]{\ifmmode
	   \oldmsb{#1}%
	 \else\strut\raise-.54ex\hbox{\scriptsize #1}\fi}
\newcommand\msp[1]{\ifmmode
	   \oldmsp{#1}
	 \else \math{\oldmsp{#1}}\fi}
\newcommand\msb[1]{\ifmmode
	   \oldmsb{#1}
	 \else \math{\oldmsb{#1}}\fi}
\newcommand\supon{\catcode`\^=7}
\newcommand\supoff{\catcode`\^=12}
\newcommand\subon{\catcode`\_=8}
\newcommand\suboff{\catcode`\_=12}
\newcommand\supsubon{\supon \subon}
\newcommand\supsuboff{\supoff \suboff}

\newcommand\actcharon{\catcode`\~=13}
\newcommand\actcharoff{\catcode`\~=12}

\newcommand\paramon{\catcode`\#=6}
\newcommand\paramoff{\catcode`\#=12}

\comment{And now to turn us totally on and off...}

\newcommand\reservedcharson{\commenton \mathshifton \atabon \supsubon \actcharon
	\paramon}

\newcommand\reservedcharsoff{\commentoff \mathshiftoff \ataboff
	\supsuboff \actcharoff \paramoff}

\newcommand\nojoe[1]{\reservedcharson#1\reservedcharsoff}

\catcode`@=12
\reservedcharsoff

\actcharon

\begin{document}

\title{Models of the SL9 Impacts I. Ballistic Monte-Carlo Plume}
\author{Joseph Harrington}
\affil{Department of Astronomy, Cornell University}
\affil{Ithaca, NY 14853-6801}
\email{jh@oobleck.tn.cornell.edu}
\and
\author{Drake Deming}
\affil{Planetary Systems Branch, Code 693, NASA Goddard Space Flight Center}
\affil{Greenbelt, MD 20771-0001}
\email{ddeming@pop600.gsfc.nasa.gov}

\comment{\slugcomment{
\tablebox{Revision 1.0 Submitted to {\em ApJ} Fri Oct 13 18:12:00 EDT 2000.}
\tablebox{Revision 2.0 Submitted to {\em ApJ} Tue Apr  3 17:47:27 EDT 2001.}
}}

\begin{abstract}

We model the plumes raised by impacting fragments of comet
Shoemaker-Levy 9 to calculate synthetic plume views, atmospheric
infall fluxes, and debris patterns.  Our plume is a swarm of ballistic
particles with one of several mass-velocity distributions (MVD).  The
swarm is ejected instantaneously and uniformly into a cone from its
apex.  Upon falling to the ejection altitude, particles slide with horizontal
deceleration following one of several
schemes.  The model ignores hydrodynamic and Coriolis effects.
Initial conditions come from observations of plume heights and
calculated or estimated properties of impactors.  We adjust plume
tilt, opening angle, and minimum velocity, and choose MVD and sliding
schemes, to create impact patterns that match observations.  Our best
match uses the power-law MVD from the numerical impact model of
\citeauthor{zahmsmlcsljgr}, with velocity cutoffs at 4.5 and 11.8
km/sec, cone opening angle of 75{\degrees}, cone tilt of 30{\degrees}
from vertical, and a sliding constant deceleration of 1.74
m/sec\sp{2}.  A mathematically-derived feature of
\citeauthor{zahmsmlcsljgr}'s published cumulative MVD is a thin shell
of mass at the maximum velocity, corresponding to the former
atmospheric shock front.  This vanguard contains 22% of the mass and
45% of the energy of the plume, and accounts for several
previously-unexplained observations, including the large, expanding
ring seen at 3.2 {\microns} by \citeauthor{mcgnacosljic} and the
``third precursors'' and ``flare'' seen near 300 and 1000 sec,
respectively, in the infrared lightcurves.  We present synthetic views
of the plumes in flight and after landing and derive infall fluxes of
mass, energy, and vertical momentum as a function of time and position
on the surface.  These fluxes initialize a radiative-hydrodynamic
atmosphere model (Paper II of this
series\comment{\citeauthor{demhslrhaic}}) that calculates the thermal
and dynamical response of the atmosphere and produces synthetic
lightcurves.

\end{abstract} 

\section{Introduction}
\label{intro}

The impacts of comet Shoemaker-Levy 9 (SL9) fragments into Jupiter's
atmosphere were perhaps the most observed events in the history of
professional astronomy.  Yet, numerous questions about the impacts and
the striking patterns they left in the Jovian atmosphere remain
unanswered.  This paper series seeks to explain as many of the
observed SL9 impact phenomena as possible using simple, consistent
physical models.  Modeling the events is challenging because of the
extreme ranges of size and energy involved.  Temperatures cover four
orders of magnitude.  Velocities cover six orders, bracketing the
atmospheric sound speed.  The volume of the affected region starts at
\sim1 km\sp{3} but material and/or heat spreads radially in a few
hours to more
than 18,000 km from the impact site \citep{mcgnacosljic} after rising
at least 3,000 km in some impacts \citep{jescbhslplic}.  It is
computationally intractable on today's
computers to model the events in a single code that covers all of the
relevant physics and chemistry.  Modelers have therefore divided the
event into phases.  The dominant physics is different in each phase,
as are the relevant materials, length scales, speeds, and durations
(see Table \ref{phases}).  So far, only \citet{ahrtooisljgrl} treat
more than a single event phase in one model.

However, several modelers of the entry and entry response phases
\citep{crabtrisljsw,zahmcjslic} linked their two models by
initializing the second phase with the results of the first.  Our
approach is similar, but for the plume flight and landing response
phases.  This paper describes our ballistic Monte-Carlo plume model.
We initialize the model with the published final velocity distribution
of \citet[hereafter ZM95]{zahmsmlcsljgr}.  It calculates the density
of the flying plume and the mass, energy, and vertical momentum fluxes
on the atmosphere, both as functions of time and position in the
relevant domain.  From this information we produce synthetic views of
the plumes in flight and of the post-impact atmospheric appearance.
We vary the free geometric parameters until the views it produces most
closely match the observations.

The plume model also handles post-re-entry sliding of material in a
parameterized fashion.  This is a ballistic model and sliding of
material upon entry into the atmosphere is certainly dominated by
hydrodynamic processes.  Nevertheless, we have included several
schemes for sliding as first-cut means of producing views of the
impact patterns.  This lets us iteratively set the model's free
geometric parameters so that they produce the most realistic final
picture, and follows our goal of parameterizing the plume's behavior
so that more realistic models can later refine the agreement with
observations without exploring a large parameter phase space.

The ballistic plume lands on the upper boundary of our
radiative-hydrodynamic atmosphere model \citep[hereafter Paper
II]{demhslrhaic}, which produces lightcurves and atmospheric
temperature and pressure fields.  We thus synthesize a composite model
whose important physics and scales adapt to the changing phases of the
events, while preserving the physical state as much as possible from
impact through landing response.  By projecting the entry-response
models' final conditions forward to the landing response period, we
can test entry-response-phase models by comparing the resulting
synthetic observables to the actual data.

The next section summarizes the measurements regarding motion and
position of material, prior models, and the community consensus
explanations for phenomena, where they exist.  We then describe our
plume model in detail, investigate the effects of varying the free
parameters, and discuss implications for modeling of subsequent
phases.  We conclude with a summary and an outline of future work.

\section{Event Phases, Observations, and Models}
\label{obs}

We summarize here the impact phenomena and previous modeling efforts,
organized by phase.  The number of SL9 papers precludes exhaustive
references.  Table \ref{phases} summarizes the event phases and
characteristics.

\end{multicols}
\atabon\begin{deluxetable}{lcccc}
\tablecaption{\label{phases} SL9 Event Phases}
\tablewidth{0pt}
\tablehead{
\colhead{Phase}    &
\colhead{Duration} &
\colhead{\tablebox{Scale \\ \hfil(km)\hfil}}    &
\colhead{\tablebox{\hfil\mvmax\hfil \\ \hfil(km/s)\hfil}}   &
\colhead{\tablebox{Temperature \\ \hfil (K)\hfil}}}
\startdata
Entry            & few min       & few\by 100 &  60 & 100--40,000+ \\
Entry response   & few\by 10 min & few\by 100 &  60 & 100--40,000 \\
Plume flight     & 20 min        &     20,000 &  12 & 8,000--10 \\
Landing response & few hours     &     20,000 &  12 & 100--3500 \\
Dissipation      & years         &     global & 0.2 & \sim100 \\
\enddata
\end{deluxetable}\ataboff
\begin{multicols}{2}
\comment{\placetable{phases}}

Entry phase models handle the interaction of a 60 km/sec cometary
fragment travelling through the stationary atmosphere.
\citet{choyoicsliau} accurately calculate fragment velocities from
pre-impact observations and the locations of the impact sites, but the
size, mass, and composition of the fragments are still debated
\citep[Paper II]{cardewtseslgic}.  A comet fragment disrupts as it
falls, vaporizing entirely and depositing most
of its kinetic energy near its terminal atmospheric depth.
Hypervelocity hydrodynamics, friction, radiation from the shock, and
material ablation dominate.  Because of the high resolution
needed, complex physics that includes all phases of matter and
transitions between them, and the long interval relative to the
timescale of dynamical interest, all models of this phase but one use
either analytical or 2D approximations.  \citet{cardewtseslgic},
\citet{crabtrisljsw}, and \citet{zahmcjslic} calculate peak
temperatures of 30,000--40,000 K.  One must resolve the fragment
finely, with grid scales of \sim1 meter \citep{zahmcjslic}.  Very
little was directly observed from the entry phase, except for a small
(as viewed from Earth) ``first precursor'' bolide flash reflected from
trailing cometary material \citep{bosctrnmslfiogrl}.

Three groups published gridded 2D numerical models, reporting on
different aspects of the same models in separate papers.  All adjusted
the atmospheric profile to compensate for the 45{\degree} inclination
of the impact vector and used cylindrical symmetry about the channel
axis.  The Sandia group
\citep{bosctrnmslfiogrl,craslmfpfeiau,crabtrisljsw} used their
laboratory's CTH and PCTH codes.  \citet{zahmcjslic,zahmsmlcsljgr},
\citet{macefmvoiau}, and \citet{zahplumesiau} used ZEUS, which we also
use in Paper II.  \citet{shohrnslpcgrl} used MAZ, but the the model was
not fully developed at the time of Dr.\ Shoemaker's death.  CTH and
MAZ were developed to model nuclear explosions, while ZEUS was
developed for astrophysics, so the physical conditions of the impacts
are not extreme for these codes.  The smoothed-particle hydrodynamics
(SPH) model of \citet{ahrtooisljgrl,ahrtorsisljgrl},
\citet{takoaslsijic}, and \citet{takaslsopic} is unique of those
mentioned here as it handles both the entry and entry response phases
in a single, 3D model.

\citet{zahmcjslic} questioned whether the SPH model could resolve the
instabilities responsible for breaking up a fragment.
\citet{takaslsopic} addressed these concerns insufficiently, in our
view, by presenting additional models that did not meet the resolution
criteria of \citet{zahmcjslic}.  However, they also point out the
shortcomings of 2D models, and called for open code comparison that to
our knowledge did not take place.  Since the models differ by an order
of magnitude in their prediction of the impact depth, we do not
consider any of the entry models to be definitive at this point.
Since their results strongly affect the details of subsequent phases,
we encourage these groups or others to continue detailed entry
modeling.  We urge those who do so to publish their final plume
geometry and mass-velocity distribution and to deposit digital final
model grids with the NASA Planetary Data System to allow models of
later phases.

The entry response phase is somewhat arbitrarily separated from the
entry phase by the need to cover a different spatial scale.  Here, the
atmosphere responds to the new energy, momentum, and material it
receives from the fragment.  A shock propagates down and outward.
Heated gas travels back up the entry channel, forming a plume that
leaves the atmosphere at speeds of at least 10 km/sec.  The atmosphere
continues to adjust to these events.  Shock physics and
non-hydrostatic hydrodynamics dominate.  A set of expanding rings left
the impact region at 454 m/s and 180--350 m/s
(\citeauthyear{hambiohapslsci}, hereafter HAM95;
\citeauthyear{harlbddjaslnat}; \citeauthyear{ingkwslnat};
\citeauthyear{walbswdsljaic}), but, as predicted by
\citet{harlbddjaslnat}, no other significant impact-related dynamical
disturbances were observed in the atmosphere.  The four entry modelers
continue in this phase with their respective codes.  Sandia and
Shoemaker switch to 3D and re-orient the entry channel to
45{\degrees}.  The SPH group continued with the same computational
grid as before, while the others all change resolutions.

\begin{figure*}
\plotone{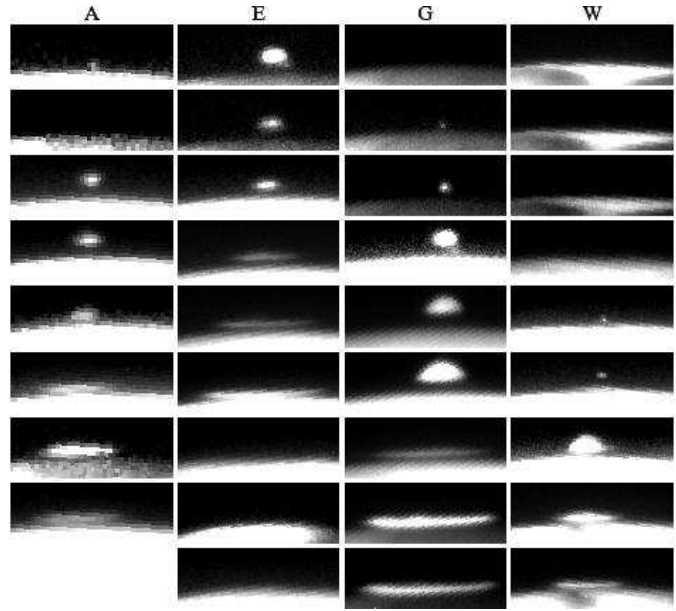}
\caption{\label{hstplume}
HST images plumes from the A, E, G, and W impacts, respectively, seen
on Jupiter's limb.  The plumes continued to slide after they had fully
collapsed.  Note emission from the hot ejection tube in the first E
and fourth G images.
}
\end{figure*}
\comment{\placefigure{hstplume}}

\begin{figure*}
\plotone{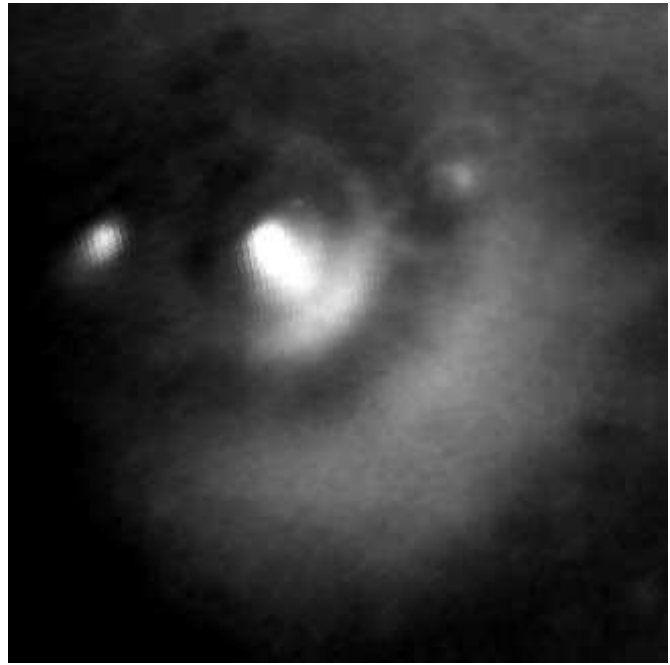}
\caption{\label{hstgd}
The G impact site, orthographic projection.  Debris is bright in this
889 nm methane band HST image.  The impact site is at the center of
the complete ring, which is a propagating wave.  It is located just
inside the northwest portion of the streak region.  The inner edge of
the crescent slid less near the axis of symmetry than away from it,
whereas the outer edge slid more near the axis.  The tiny D impact
streak is on the left.
}
\end{figure*}
\comment{\placefigure{hstgd}}

Plume flight is a
phase in its own right.  The Hubble Space Telescope observed four
plumes (impacts A, E, G, and W, HAM95).  \citet{jescbhslplic}
calculated maximum 
altitudes of 2,300--3,100 km for material visible in these plumes.
This region is so rarefied that
hydrodynamics have minimal effect in the 20 minutes or so of flight,
and ballistics dominate (ZM95).  Initially, a plume is so hot it emits
in the visible: the Galileo Photopolarimeter Radiometer and
Ultraviolet Spectrometer both measured emission at 8,000 K
\citep{cardewtseslgic}.  This emission caused the ``second
precursors'' in the observed lightcurves \citep{bosctrnmslfiogrl}.
However, adiabatic expansion quickly cools the ejected gas to a few K,
and the lightcurves are again quiet.  \citet{shohrnslpcgrl} suggest a
maximum flight time of only 10 minutes, but the observations show
otherwise.  The Sandia group ran their entry-response model through
the plume phase, but stopped short of presenting the atmospheric
re-entry boundary conditions or detailed impact images, as we do here.
However, their code is capable of modeling the hydrodynamics relevant
in the early part of this phase, which we cannot.

The present work, \citet{paniepicslic}, and ZM95 all simplify the
physics to ballistics at this stage.  ZM95 justifies this
simplification analytically.  Intuitively, the pressures in the early
stages of expansion into a vacuum are much larger than any others that
will be encountered.  The initial accelerations will dominate
subsequent ones except those that act over long periods of time.
After the initial expansion, the only consistent force is gravity, as
pressure drops with volume and volume increases as the cube of time
(faster, in the first seconds when hydrodynamics are relevant).  The
purposes of the three models diverge at this point.
ZM95's Monte-Carlo ballistic model is a vertical 2D sheet, and was
used to study lightcurves and chemistry.  \citet{paniepicslic}
constrain plume sizes by parametric simulation of sliding under
Coriolis influence.  We calculate the boundary conditions for the
atmospheric response.  The present work, \citet{paniepicslic}, the
Sandia group, and the SPH group all calculate synthetic impact-site
views.

In the landing response phase, falling plume material compressed
itself and the ionosphere and upper stratosphere until they radiated,
causing the entirely forseeable and quite unpredicted ``main event''
in the lightcurves \citep{zahplumesiau}.  Shocks travelled up through
the infalling plume material and down into the atmosphere.  The
infalling material deposited its vertical momentum, but continued its
horizontal motion for many minutes, sliding on the atmosphere
\citep[HAM95,][]{jescbhslplic}.
At wavelengths near 0.9 {\microns}, a brief ``flare'' appeared and
vanished 1000 seconds after impact
\citep{schbksilsljemp,fitacoijslmnras,ortomslhic}, coincident with
spectral observations of hot CO \citep{meaceslw}.  In some lightcurves
the flare briefly outshone the main event.  The lightcurves then fell
nearly as quickly as they rose, but not completely back to their
initial level.  They oscillated or ``bounced'' several times with a
10-minute period \citep{nicghslrlcgrl}, then decayed for tens of
minutes, depending on the wavelength observed.  At 1.25 and 2 hours
after impact, \citet{mcgnacosljic} observed a ring with radii of
14,000 and 18,000 km, respectively, giving a mean expansion rate of
nearly 1.5 km/s in this time period.  We call this feature
``McGregor's Ring.''

Eventually, material velocities fell below the sound and then the wind
speeds.  An unknown process generated a dark material that remained
high in the atmosphere in the shape of a large crescent.  This debris
pattern's outer edge is \sim13,000 km from the largest impact sites.
Its inner edge is located at \sim6,000 km.  The
symmetry axis that splits the crescent and contains the impact site is
rotated 14{\degrees}-21{\degrees} from the surface track of the
incoming fragment.  \citet{paniepicslic} and \citet{jescbhslplic} model
this as a Coriolis effect
during plume flight and sliding.  The material is bright in methane
bands (HAM95), indicating a depth of just a few mbar
\citep{molmmadccslic}.  There is more material near the impact site
itself, and \citep{walbswdsljaic} claim the entry response rings are
also made visible by this material.  The largest impacts have
rays pointing 1000--2000 km downrange from the impact site.  These may
be due to Rayleigh-Taylor instabilities in the earliest
(pre-ballistic) plumes (HAM95).  The debris crescent's composition is
uncertain \citep{weskfsldebsci} and it is still referred to by workers
in the field as ``the brown gunk.''

In the dissipation phase, material spreads with the winds, covering
all longitudes in a few weeks and the southern mid-latitudes over
several years, eventually fading from view
\citep{bangssjsaic,simbjtfwfllsic}.

HAM95 observed the plume material sliding on the atmosphere (see Fig.\
\ref{hstplume}), an interpretation corroborated by the crescent
locations.  The G impact plume rose to \sim3,100 km
\citep{jescbhslplic}, but the outer crescent edge for that impact is
\sim13,000 km from the impact site, more than twice as far as a
ballistic object can fly
under gravity, given this maximum height and assuming no bias in speed
with ejection direction.

\section{Ballistic Model}
\label{ballmodel}

Ours is a Monte-Carlo ballistic model with parameterized sliding after
reentry.  At present we ignore Coriolis effects, planetary curvature,
hydrodynamics, and thermodynamics.  The model runs in two modes,
called ``fly'' and ``land''.  Fly mode calculates the plume mass
density at specific post-impact times in the volume of space above the
impact site, creating views of the plumes in
flight.  Land mode calculates the plume re-entry mass, energy, and
vertical momentum fluxes as a function of time and position on the
surface.  This mode both creates synthetic views of the final impact
site and initializes our atmospheric
response model (see Paper II).

We divide a plume into discrete, independent mass elements whose
behavior is the same in both modes.  Each element gets a random speed
and direction.  The model calculates the flight path and the time and
location at which the element returns to its initial level.  Upon
landing, the element deposits its vertical momentum and slides
horizontally until it stops, where it deposits its mass and energy.
Both modes place a virtual grid of bins in the area of interest, and
increment the appropriate bin when an element arrives.
When the model has run the specified number of elements,
it scales the bin values by volume and particle mass.  In
both modes one may choose either a rectangular or a spherical/polar
grid, and may choose the minimum, maximum, and bin interval values
independently on each axis.  The impact site is the origin.  In fly
mode, the grid has three spatial dimensions, is evaluated at specific
post-impact times, and accumulates only mass.  In land mode, the grid
is on the surface and thus has two dimensions spatially plus one
temporally.  It accumulates mass, energy, and vertical momentum.

\begin{figure*}
\centerline{
\includegraphics[clip,width=0.6\columnwidth,angle=270]{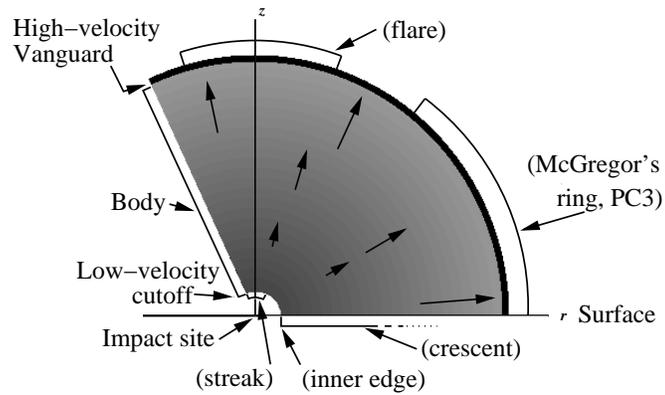}}
\caption[Plume cutaway diagram.]{\label{plumediag}
Plume cutaway diagram.  All mass is instantaneously ejected up into a
cone from the impact site, which is located at its (downward-pointing)
apex.  Vectors in the plume body indicate initial velocities.  This side
view shows a plume's principal parts, the surface effects to which
they give rise, and (schematically) the mass-velocity distribution.
The instantaneous ejection implies that each plume element ``sees''
all other elements moving away with velocities proportional to their
distances.  There is thus no friction.
}
\end{figure*}

The velocity distribution separates into independent direction and
speed components.  The ejection directions are uniformly distributed
over the solid angle within a specified angle of the ejection channel
axis.  Figure \ref{plumediag} shows a cutaway view of this cone-like
geometry.

\comment{\placefigure{plumediag}}

\begin{figure*}
\plotone{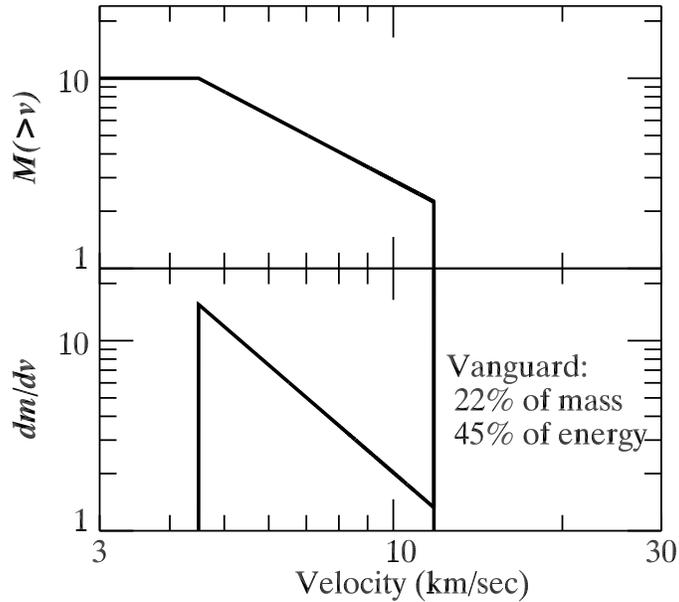}
\caption{\label{zmmvd}
Top: Cumulative velocity distribution of ZM95, as modified for our
model.  Bottom: Differential velocity distribution used in the present
model, with outer mass shell (derived in text).  Cumulative and
differential mass are in the same arbitrary units, which scale with
\math{k}.
}
\end{figure*}
\comment{\placefigure{zmmvd}}

The model has two different MVDs.  The first is the uniform
distribution between two cutoff velocities.  The second is an
analytical model for the distribution that fits the numerical
experiments of ZM95 very well (see their Fig.\ 3 and our Fig.\
\ref{zmmvd}).  We chose
the ZM95 distribution because it is the only one published in a form
that is conducive to further modelling; most others only published
synthetic images of specific energy or temperature and none published
plots of mass {\em vs}.\ velocity.  The ZM95 distribution also closely
follows theoretical distributions for hypervelocity impact ejecta
(ZM95 reviews theory and its applicability to this problem).  Further,
ZM95 developed the analytical approximation to their numerical
distribution themselves, making clear their interpretation of their
own numerical data:

\begin{equation}\atabon
\label{cummass}
M(>v) = \left\{ \begin{array}{r@{\quad:\quad}l}
k v\sp{-\alpha} & v \le \vmax\\
0               & v >   \vmax
\end{array}\right.,
\ataboff\end{equation}
where \math{M(>v)} is the mass moving at speeds greater than \math{v},
\math{k} is a constant (see below), {\mvmax} is the cutoff velocity
seen in Fig.\ \ref{zmmvd}, and \math{\alpha} is a constant between 1
and 2, constrained to that range by conservation considerations.  The
data in ZM95's Fig.\ 3 give \math{\alpha=1.55}.  This model approaches
infinite mass at low velocity, but the energy is finite.

Cumulative distributions can make it difficult to visualize how the
mass is distributed: just how much more mass exists between 1 and 2
km/s than between 9 and 10 km/s?  A differential distribution would be
more helpful, and is necessary for modeling.  Taking the derivative
piecewise yields:

\begin{equation}\atabon
\label{diffmassfirst}
\frac{\mdiff m}{\mdiff v} = \left\{ \begin{array}{r@{\quad:\quad}l}
k \alpha v\sp{-\alpha-1} & v < \vmax \\
0                        & v > \vmax
\end{array}\right..
\ataboff\end{equation}

There is no leading minus sign because the sense of accumulation was
\math{M(>v)} rather than \math{M(<v)} before taking the derivative,
which changes the sign of the right side of Eq.\ \ref{diffmassfirst}.
One discovers something peculiar about this distribution.  If
\math{M(>v) = 0} at speeds faster than {\mvmax}, how can it
discontinuously be positive at \math{v < \vmax}?  The discontinuity in
Eq.\ \ref{cummass} yields a delta function in the derivative of size
\math{M(>\vmax)}, which at first appears unphysical.  One might be
tempted to dispose of such a mathematical quirk, but ZM95's Fig.\ 3
comes from the final data of a numerical SL9 impact model.  It is not
the result of computational peculiarities of that model and it is in
line with a well-established theory.  Their plotted line does not cut
off abruptly, but rather drops off with a very steep slope
(\math{\alpha \approx 6}), which still yields a spike in the
derivative that integrates to \math{M(>\vmax)}.  The spike appears as
a dense outer shell of gas in Fig.\ 4 of
\citet{ahrtooisljgrl}, which
shows the effect to be greater for larger fragments.  We
hypothesize that this shell is the remnant of the shock wave and
compressed upper atmosphere that preceeded the main body of the plume
into space.  We call this spike of mass the plume ``vanguard,'' since
it is the first plume material to reach any given point and since its
effect upon landing is similar to that of its namesake.

As we discuss below, the vanguard contains 22% of the mass and 45% of
the energy in the model plumes whose final appearance best matches the
HST data.  We find below that it explains McGregor's ring, a finding
confirmed more rigorously in Paper II.  That paper further finds that
it produces the
``third precursor'' (PC3) and flare features of the lightcurves
\citep{fitacoijslmnras}.  McGregor's ring and the flare are
absent in runs without a vanguard.  We also note that the vanguard is
similar to the single mass shell found by \citet{paniepicslic} to
match their data well.  Finally, we point out that a density
enhancement is often seen immediately behind shocks (e.g., see the
shock tube test of Paper II) and would therefore not be a surprising
feature of the initial plume expansion.

To create post-impact views with a debris crescent (see below), we
also introduce a minimum velocity (\mvmin), which may be zero for the
flat distribution but must be positive for the power-law distribution.
One physical justification for a minimum velocity cutoff is the
``pinching off'' at the 1-bar level suggested by
\citet{bosctrnmslfiogrl} to explain why the plumes all rose to the
same height.  Another, suggested by \citet{zahmlfscwsljgr}, is that
carbonaceous grains, which may be the brown gunk, can be formed
by material impacting faster than 4.5 km/s.
This cutoff does not introduce any further complications into
the differential distribution used in the model.  We call the final
distribution ``cumulative power law with cutoff'', or CPC:

\begin{equation}\atabon
\label{diffmass}
\begin{array}{lr}
\lefteqn{\frac{\mdiff m}{\mdiff v} =} & \\*[2ex]
\left\{\begin{array}{r@{\;:\;}l}
  k (\alpha v\sp{-\alpha-1} + \vmax\sp{-\alpha}\delta(v-\vmax)) 
		 & \vmin \le v \le \vmax\\
  0 & \mbox{otherwise.}
 \end{array}\right.
\end{array}
\ataboff\end{equation}
\comment{ & \mbox{``body'' + ``vanguard''} \\}

To assign each particle a velocity, we first calculate
\math{\mvan/\mplume}, where \math{\mvan = M(>\vmax)} and \math{\mplume
= M(>\vmin)}, and assign {\mvmax} to that fraction of the particles.
For the remaining particles, we integrate Eq.\ \ref{diffmass} from
{\mvmin} to \math{v'} and divide by the total mass to give an
expression for \math{M(<v')} with our particular {\mvmin}, and solve
for \math{v'}.  This gives the velocity probability distribution in
terms of a uniformly distributed random variable, \math{r}, whose range is 0--1:

\begin{equation}\atabon
\label{velint}
v(r) = 
\left(\vmin\sp{-\alpha}
- r(\vmin\sp{-\alpha} - \vmax\sp{-\alpha}) 
\right)\sp{-\frac{1}{\alpha}}
\ataboff\end{equation}
A random number generator provides
uniformly-distributed numbers, which we scale to the range 0--1.

We randomly assign a particle's azimuth and deflection (\math{\theta})
around an initially vertical axis, then convert to rectangular
coordinates and rotate the direction vector to the axis inclination
and azimuth specified for the plume.  The distribution of azimuth at a
given axis deflection is inherently uniform in polar coordinates, so
we simply rescale the random numbers to the range 0--\math{2\pi}.  The
histogram of \math{\theta} is proportional to the
circumference of circles of constant \math{\theta}, and we follow a
similar procedure to the derivation of Eq.\ \ref{velint} to create an
appropriate function of \math{r} and the opening angle \math{\theta'}:

\begin{equation}\atabon
\label{colatredist}
\theta(r) = \cos\sp{-1}\left[1-r(1-\cos \theta')\right].
\ataboff\end{equation}

The constant \math{k} embodies the bulk impact physics other than the
MVD and geometry.  This number characterizes the density of the plume
and the amount of entrained Jovian air.  It scales with the impactor
mass (\math{m\sb{i}}), velocity (\math{v\sb{i}}), and the fraction of
impactor kinetic energy going into the plume (\math{\eta}).  Following
ZM95 but including the mass in the vanguard, we set the total
available energy equal to that in the plume:

\begin{equation}\atabon
\label{enint}
\frac{\eta m\msb{i}v\msb{i}\msp{2}}{2}
= \int\msb{\vmin}\msp{\vmax}\frac{\mdiff m}{\mdiff v}\frac{v\msp{2}}{2}\mdiff v.
\ataboff\end{equation}
Substituting \math{\frac{\mdiff m}{\mdiff v}} from Eq.\ \ref{diffmass}
and solving for \math{k} gives:

\begin{equation}\atabon
\label{konst}
k = \frac{\eta m\msb{i}v\msb{i}\msp{2}(2-\alpha)}{2\vmax\sp{2-\alpha}-\alpha\vmin\sp{2-\alpha}}.
\ataboff\end{equation}
For comparison to ZM95, the cumulative mass with \mvmin=0 is

\begin{equation}\atabon
\label{kzcompcummass}
M(>v) = \frac{2-\alpha}{2}\eta m\msb{i}
\left(\frac{v\msb{i}}{\vmax}\right)\sp{2}
\left(\frac{\vmax}{v}\right)\sp{\alpha},
\ataboff\end{equation}
and \math{\mvan} is just this with \math{v=\vmax}.  This differs from
ZM95's equation 9 only in the replacement of \math{\alpha} with 2 in
the first denominator.  The expression for vanguard energy
\math{e\sb{\mbox{\scriptsize van}}} under the \mvmin=0 assumption is

\begin{equation}\atabon
\label{kzcummass}
e\sb{\mbox{\scriptsize van}}
= \frac{\eta m\msb{i}v\msb{i}\msp{2}}{2}\left( 1-\frac{\alpha}{2}\right).
\ataboff\end{equation}

Only \math{v\sb{i}} is observationally constrained, at about 60 km/s.
The product \math{\eta m\sb{i}} is not directly constrained by this
model, nor do any results presented in this paper depend on it.  ZM95
state that their impact models are consistent with values in the range
\math{0.3 \le\eta\le 0.45}, but in their discussion they allow values
as low as 0.01.  \citet{bezgklissltric} estimate \math{\eta>0.2} for
the L impact.  We use \math{\eta}=0.3.  The mass \math{m\sb{i}} is one
of the holy grails of SL9 post-impact work \citep{cardewtseslgic}.
Our nominal value is 1.4\tttt{14} g, which is justified in Paper II
based on lightcurve intensity levels and ZM95's \math{\eta}.  
Plume mass density and the plume re-entry fluxes of
mass, energy, and momentum are all linearly proportional to \math{k},
which is in turn independent of \math{v}.  We thus compute the entire
model using counting bins that ``catch'' particles of unit mass, and
multiply the bins by a scale factor that includes \math{k} at the end
of the computation.  This allows flexibility in choosing the value of
\math{\eta m\sb{i}}, as the entire grid can again be multiplied by a
scale factor until models based on it produce synthetic observables
most in line with observations.  This can be done without rerunning
the model.

For consistency with ZM95, we evaluate \math{k} under the assumption
of an MVD running from 0 to \math{\vmax}.  This only affects the total
mass, which we take as unconstrained and will later fit.  The relative
size of the vanguard and the rest of the distribution remains the
same.

Upon landing and in land mode only, the particles deposit their
vertical momentum at the azimuth, radial distance (\math{r\sb{l}}),
and time (\math{t\sb{l}}) of landing, and slide with initial
horizontal velocity \math{v\sb{h}} following one of three schemes: no
sliding, constant time, and constant deceleration.  The sliding
schemes calculate the distance (\math{r\sb{s}}) and duration
(\math{t\sb{s}}) of sliding, and deposit particle mass and kinetic
energy at \math{r\sb{l} + r\sb{s}}, \math{t\sb{l} + t\sb{s}}, and the
landing azimuth.  We carry the energy along because we expect most of
it to convert to heat, which slides with the particle.

In the ``no sliding'' scheme, all quantities are deposited at the
landing location and time.  ``Constant-time'' sliding was inspired by
conference discussions reporting that sliding occurred for about
\math{t\sb{s}}=20 minutes after impact, based on a quick look at the
HST data.  Particles travel \math{r\sb{s} = t\sb{s}v\sb{h}}.  With
constant deceleration \math{f}, \math{r\sb{s} = v\sb{h}\sp{2}/2f} and
\math{t\sb{s} = v\sb{h}/f}.

\section{Model Runs and Discussion}
\label{rundisc}

The previous section makes it apparent that the model has a large
number of adjustable parameters.  However, direct observations
constrain many of their values and the input velocity distribution
(based on prior models) constrains several others.  Most of the
remainder have identifiable, independent effects on the debris pattern
views, and this independence reduces the volume of reasonable phase
space dramatically.

\end{multicols}
\atabon\begin{deluxetable}{l@{\extracolsep{0.25em}}cr@{\extracolsep{0.25em}}l@{\extracolsep{0.25em}}l}
\tablecaption{\label{paramtab} Model Parameters}
\tablewidth{0pt}
\tablehead{
\colhead{Description} &
\colhead{Symbol} &
\colhead{Value} &
\colhead{Units} &
\colhead{Constraint}}
\startdata
gravity                                  & \math{g}         & 23.25         & m/s\sp{2}   & orbits, day length, latitude \\
impactor velocity                        & \math{v\sb{i}}   & 60            & km/s        & \citet{choyoicsliau} \\
impactor mass                            & \math{m\sb{i}}   & 1.4\tttt{14}  & g           & paper II \\
plume energy fraction                    & \math{\eta}      & 0.3           &             & ZM95 \\
opening angle                            & \math{\theta}    & 75            & \degrees    & fit \\
axis azimuth (N \math{\rightarrow} E)    &                  & 145           & \degrees    & HAM95 \\
axis tilt from normal                    &                  & 30            & \degrees    & fit \\
mass-velocity distribution               & \multicolumn{2}{r}{CPC, flat}    &             & fit, ZM95 \\
\hspace{2em} mass-velocity power law             & \math{\alpha}    & 1.55          &             & fit, ZM95 \\
low velocity cutoff                      & \mvmin           & 4.5           & km/s        & fit, \citet{carwsstsasleso} \\
max particle velocity                    & \mvmax           & 11.81         & km/s        & HAM95 \\
sliding scheme                           & \multicolumn{2}{r}{const.\ deceleration} &     & fit \\
\hspace{2em} sliding time constant               &                  & 0             & s           & fit \\
\hspace{2em} deceleration	                 &                  & 1.74          & m/s\sp{2}   & fit  \\
\\
observed first re-entry time             & \math{t\sb{min}} & 350           & s           & \citet{carwsstsasleso} \\
observed max plume altitude              & \math{z\sb{max}} & 3,000         & km          & HAM95 \\
\enddata
\end{deluxetable}\ataboff
\begin{multicols}{2}
\comment{\placetable{paramtab}}

Table \ref{paramtab} summarizes the variable parameters and sources of
observational constraint and presents the values in our nominal
plume.  The values of \math{m\sb{i}} and \math{\eta} are very
uncertain but do not affect the appearance of the debris-field views
in this model.  In practice, the final step of the model
divides all particle counting bin values by the total number of
particles and multiplies by \math{\eta m\sb{i}}.  However, in Paper II
we discuss lightcurves, where the total mass matters quite a bit.
\citet{paniepicslic} constrain plume masses directly with their model
of Coriolis-modified sliding (but see below).  A direct estimate of
plume mass from the HST debris images might be possible, but only
after the determination of the composition, optical properties, and
means of forming the brown gunk that makes the debris field visible.
This is one of the principal unsolved puzzles of the SL9 impacts
\citep{weskfsldebsci}.

The ejection altitude is the level at which ballistics dominates
hydrodynamics in the emerging plume.  To well within the 800-km range
in the four measured plume heights \citep{jescbhslplic}, this is about
the level at which re-entering material will start to re-encounter
significant hydrodynamic forces.  Estimates of the terminal depth of
the incoming fragment vary over 200--300 km
\citep{borsrstbslic,craslmfpfeiau,macefmvoiau}.  Since the plume
travels up the path of the incoming fragment, the early development of
the entry channel and details of fragment breakup will affect the
ejection altitude.  However, these effects are small compared to the
range in plume heights.

The rest of this section presents the relationship between parameters
and observations, showing what happens to the debris-pattern views
when we change parameter values.  Each of the synthetic view panels in
the figures presents 100 million Monte-Carlo particles cast onto a
polar computational grid.  Each grid had 100 radial bins between the
impact site and the location of the most-distant debris (after
sliding), 36 azimuthal bins, and a single time bin.  Except as noted,
the views here show the surface mass density as a linearly varying
grayscale.  Because different model parameters yield vastly differing
peak densities and pattern sizes, we have scaled the intensities in
each view separately.  The total mass in any view varies only because
of changes in the minimum velocity (except when we vary the velocity
distribution), and that effect can be estimated by looking at Fig.\
\ref{zmmvd}.

\begin{figure*}
\ifthenelse{\lengthtest{\columnwidth = \textwidth}}{\epsscale{0.8}}{}
\plotone{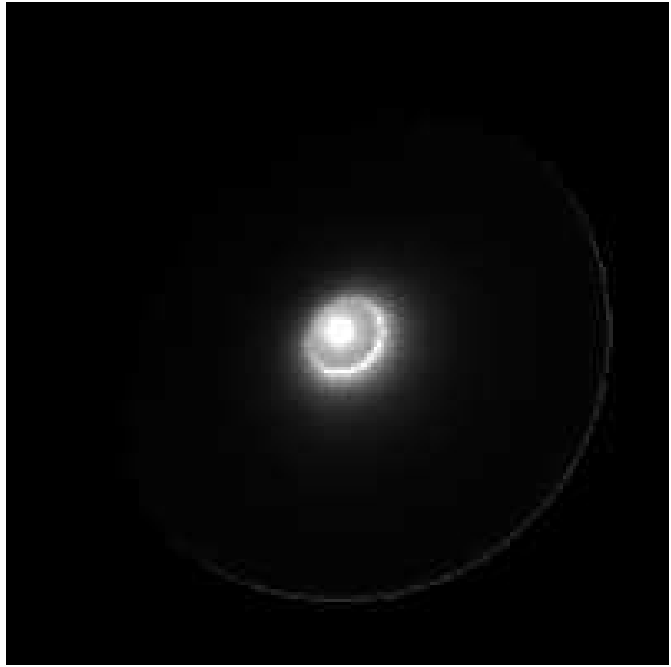}
\ifthenelse{\lengthtest{\columnwidth = \textwidth}}{\epsscale{1}}{}
\caption{\label{landideal}
Best overhead view of the impact site after all material has come to
rest.  Intensity is proportional to mass deposited.  Note that
observed impact sites show dark grains, not the plume gas.  The
formation and redistribution of grains is still not well understood,
so in these images the locations of particular features is more
important than relative intensities.  Table \ref{paramtab} gives model
parameters, which were adjusted to match the appearance of Fig.\
\ref{hstgd}.  The inside of the debris crescent has slid 6,000 km
from the impact site.  The thin exterior ring has a very large radius
(46,000 km) with our nominal sliding parameters.  We identify the ring
in our model with one observed to be expanding at nearly 1.5 km/sec 2
hours after impact by \citet{mcgnacosljic}.  The figure width is
100,000 km.
}
\end{figure*}
\comment{\placefigure{landideal}}

Figure \ref{landideal} shows the scene from above when all material
has come to rest, using the parameters in Table \ref{paramtab}.  The
low-velocity cutoff produces a sharp inner crescent edge at 6,000 km.
Vertically-ejected material landed near the impact site and stayed
near it.  This streak contains more mass than the remainder of the
debris combined, which is consistent with the data in Fig.\
\ref{hstgd}.  The crescent is narrower than in the observations, which is
likely due to the simplistic sliding applied here.  McGregor's ring
has stopped 46,000 km from the impact site.  We now adjust parameters
away from this ideal to show why these are the best values.

\begin{figure*}[tbp]
\plotone{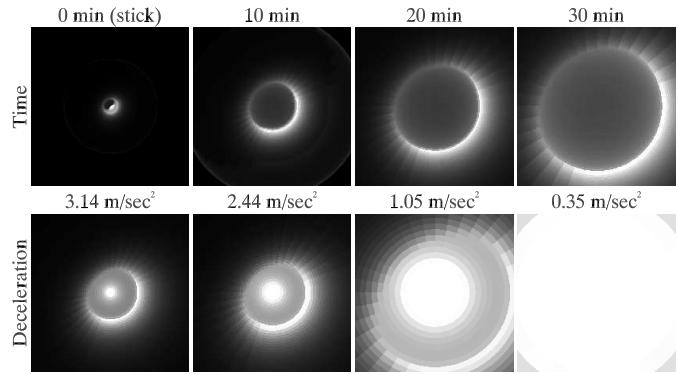}
\caption{\label{landslide}
Effects of different types of horizontal sliding.  Plot parameters are
the same as in Fig.\ \ref{landideal}.  The top-left frame has no
sliding and demonstrates the requirement for sliding of some type, as
purely ballistic motion does not deliver material nearly far enough
from the impact site.  In the remaining frames of the top row, each
particle slid at its horizontal velocity for the indicated period of
time.  This is also not realistic because it does not form a central
streak of material near the impact site.  Material that flew
vertically and thus had a very small horizontal velocity component
slides for the same time as material ejected on horizontal
trajectories, clearing the middle.  The bottom row shows sliding
with a constant deceleration that mimics work agains a constant
pressure or sliding up a hill.  This most closely matches the impact
images.  The parameters shown bracket the ideal parameter of Fig.\
\ref{landideal}.  The panel width is 20,000 km.  Vanguard material may
be outside of a given frame, but we did not use its location as a
criterion for estimating optimal parameters.
}
\end{figure*}
\comment{\placefigure{landslide}}

If we turn off sliding entirely, the top-left panel of Fig.\
\ref{landslide} shows that the material does not spread far enough.
The remaining figures in that row have constant-time sliding that
fails to produce a streak region at all.  Material ejected nearly
vertically slides for too long and leaves the central region.  Only
the bottom row's sliding with a constant deceleration does a
reasonable job, making a dense streak and a well-defined crescent.
The deceleration parameter determines where the crescent stops
sliding, and the value 1.74 m/sec\sp{2} puts the inner edge at the
observed 6,000 km.  It is coincidental that the 20-minute sliding
estimate happens also to put the inner edge near this location.

There are at least two relevant physical interpretations of constant
deceleration: work against a constant pressure and sliding up a hill.
In the hill analogy, the mass and momentum of the falling plume
material compresses the atmosphere where it lands, but closer to the
impact site there is much more material than further away.  This
creates a slope.  Debris slides in the direction away from the impact
site, trading kinetic energy for potential as it climbs the ``hill''.
Our 1.74 m/s\sp{2} deceleration corresponds to a hill pitch of
4.3{\degrees}.  In reality, the slope would vary with the
amount of material that had already fallen at a given time and how it
had moved, while our hill has a constant slope.  In the pressure
analogy, the expanding material must push against the surrounding
atmosphere to expand.  It encounters new, stationary material that it
must accelerate throughout its expansion.  In reality, the
deceleration pressure depends on the instantaneous expansion velocity,
radius, and the depth to which the plume has compressed the
atmosphere, which varies with azimuth and distance from the impact
site.  Both hill and pressure effects may play a role in stopping the
debris.

In contrast to our constant deceleration, \citet{paniepicslic} use a
per-particle force proportional to \math{v\sp{2}} that would be
appropriate for viscous drag.  Monte-Carlo particle models such as
ours and theirs ignore the fact that only a small amount of material
is involved in the turbulent boundary layer between the sliding plume
and the stationary atmosphere, and material above this interface would
continue sliding unabated until the turbulent mixing length scale was
as large as the depth of the fallen plume material, which is deep
(Paper II).  A better tool for studying horizontal expansion is a
fluid model, and we take up the question more realistically in Paper
II, initializing with unslid plumes.  Both our scheme and that of
\citet{paniepicslic} work as simple ways to simulate the appearance of
the data and to explore plume parameters, but one must be cautious if
using either model beyond that point.

\begin{figure*}[tbp]
\plotone{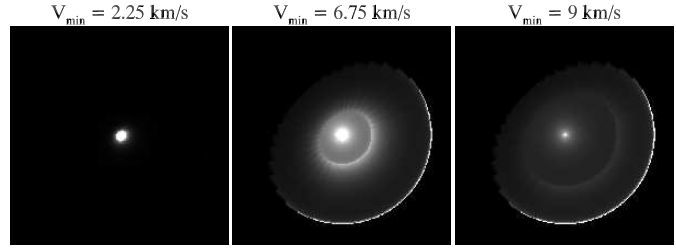}
\caption{\label{landvmin}
Effect of varying the minimum plume velocity.  Plot parameters are the
same as in Fig.\ \ref{landideal}.  The parameters shown bracket the
ideal parameter of Fig.\ \ref{landideal}.  Changing this parameter
changes the location of the inner crescent edge and the fraction of
material near the impact site.  The panel width is 100,000 km.
}
\end{figure*}
\comment{\placefigure{landvmin}}

Changes in \math{\vmin} do three things: first, they alter the
relative fraction of material in the streak, with lower \math{\vmin}
increasing the mass in the center; second, they change the location of
the inner crescent edge; and third, they change the landing time of
plume body material ejected on vertical paths.  One can adjust
\math{\vmin} and the sliding deceleration to keep the inner crescent
edge in the same location but change the relative streak and crescent
distributions.  However, doing so also changes the start time of the
lightcurve main event.

\begin{figure*}[tbp]
\plotone{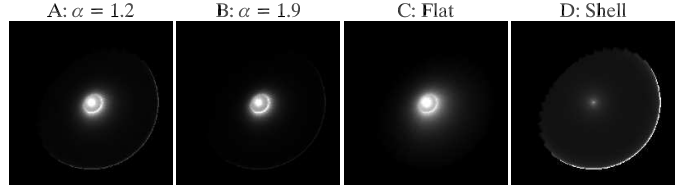}
\caption{\label{landdist}
Different velocity distributions.  Plot parameters are the same as in
Fig.\ \ref{landideal}.  A & B: ZM95 power-law with parameter
\math{\alpha} that brackets the ideal parameter of Fig.\
\ref{landideal}.  C: Flat distribution between ideal cutoff
velocities.  Its morphology is similar to that of panels A and B, but
with a sparser central region and no exterior ring.  D: Shell
distribution, essentially a delta function with velocity \math{\vmax}.
It does not resemble the data.  The panel width is 20,000 km.
}
\end{figure*}
\comment{\placefigure{landdist}}

As Fig.\ \ref{landdist} shows, changing the power law parameter
\math{\alpha} does not have a dramatic effect, but it does alter the
width of the crescent and the relative amounts of material in the
crescent and streak.  Using a flat distribution eliminates the
exterior ring feature, and would eliminate our ability to match
observations of PC3 and the flare with synthetic
lightcurves in the next impact phase and would leave McGregor's ring
unexplained.  A shell distribution (a flat distribution with
\math{\vmin} close to \math{\vmax}) eliminates the crescent.  There is
a streak because vertically-ejected material still does not slide, but
it is very small.  CPC is the only distribution that puts material in
all three places where it is observed: a central condensation, a
crescent, and McGregor's ring.  It also produces realistic lightcurves
(Paper II).  \citet{paniepicslic} conclude that a shell distribution
is more realistic than their roughly-Gaussian distribution.  Note that
our respective crescents are produced by different phenomena: theirs
is from the shell and is spread out radially by \math{v\sp{2}} drag,
ours is from CPC's low-velocity cutoff.  Our shell produces McGregor's
ring, and our constant-force sliding does not spread it very much.

\begin{figure*}[tbp]
\ifthenelse{\lengthtest{\columnwidth = \textwidth}}{\epsscale{0.65}}{}
\plotone{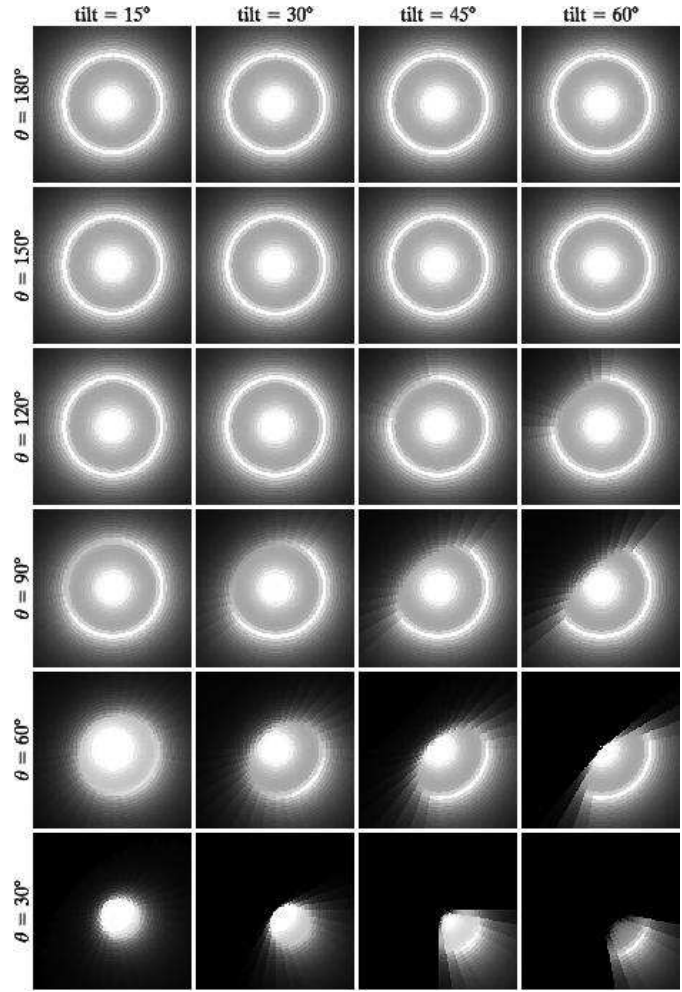}
\ifthenelse{\lengthtest{\columnwidth = \textwidth}}{\epsscale{1}}{}
\caption{\label{landgeom}
Geometry changes.  Plot parameters are the same as in Fig.\
\ref{landideal}.  Ejection axis angle from the vertical varies across
the rows as indicated, and opening angle (axis to cone side) varies
down the columns.  A 180{\degree} opening angle is a spherical
distribution.  The panel width is 20,000 km, which puts vanguard
material outside the frame.  In all cases, it is thin, circular, and
has an azimuthal mass distribution roughly in proportion to the mass interior to
it, though it tends to cover more azimuth than the interior material.
}
\end{figure*}
\comment{\placefigure{landgeom}}

Changing the cone geometry, as shown in Fig.\ \ref{landgeom}, has
dramatic effects.  The decidedly non-circular site appearance (see
Fig.\ \ref{hstgd}) requires a tilted axis, but the 45{\degree} tilt of
the impactor's entry path (tilt = 45{\degrees} column of Fig.\
\ref{landgeom}) is too much.  Our best axis is 30{\degrees} from
normal, exactly halfway between \citeauthor{craslmfpfeiau}'s
\citeyearpar{craslmfpfeiau} 20{\degrees} and
\citeauthor{takaslsopic}'s \citeyearpar{takaslsopic} 40{\degree}
values.  \citet{paniepicslic} use the entry path as the ejection axis.
The opening angle is similarly important.  A small opening angle is
like a constrained jet, whereas a large one makes the debris pattern
too circular.  We find 75{\degrees} from axis to cone edge best
reproduces the data, in good agreement with
\citeauthor{paniepicslic}'s \citeyearpar{paniepicslic} 70{\degrees}.
We do not understand how \citeauthor{takaslsopic}'s
\citeyearpar{takaslsopic} ejecta pattern, which spans well over
120{\degrees} in their Fig.\ 7, could have been generated from
non-interacting ballistic particles ejected {\pm}30{\degree} from the
bolide entry path and rotating only the stated 15{\degrees} due to
Coriolis effects.

\section{Nominal Plume Characteristics and Atmospheric Modeling
Considerations}

\begin{figure*}[tbp]
\ifthenelse{\lengthtest{\columnwidth = \textwidth}}{\epsscale{0.31}}{\epsscale{.99}}
\plotone{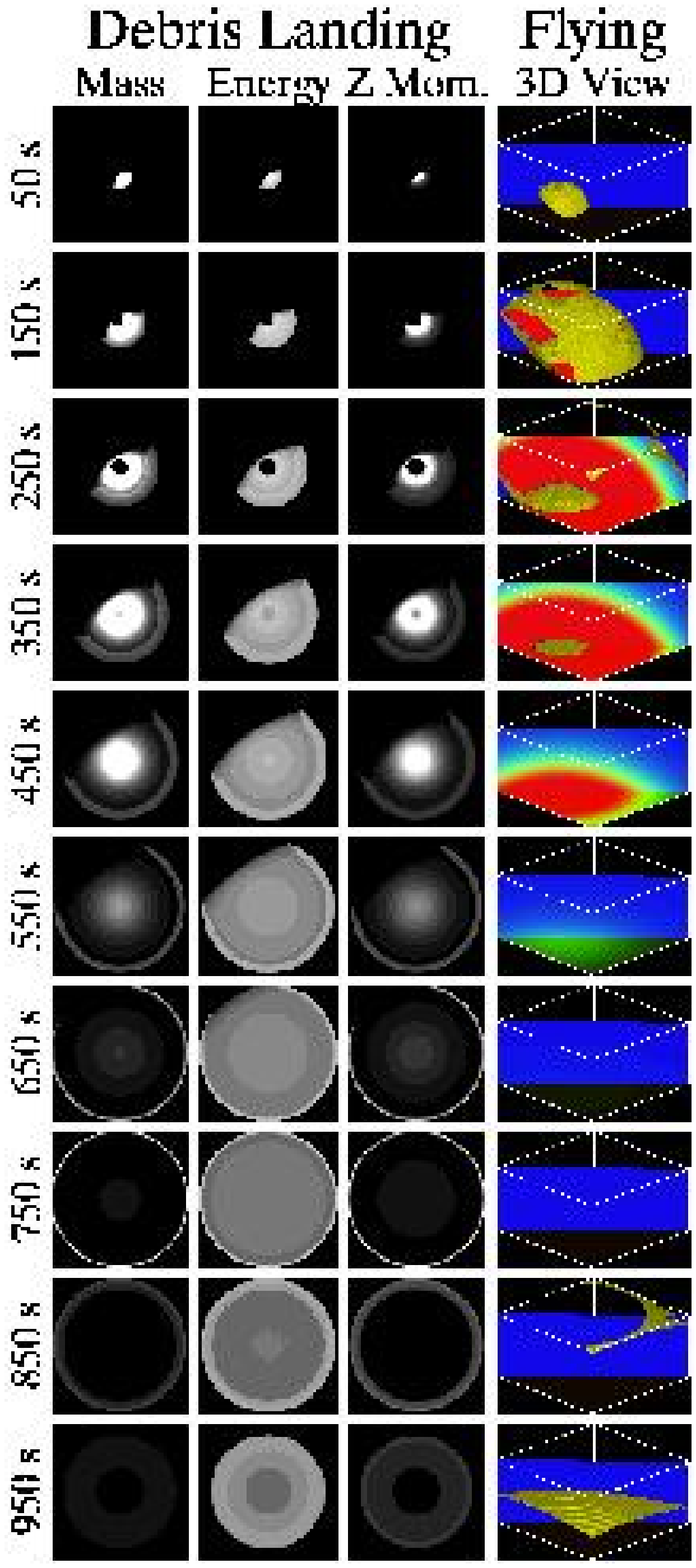}
\epsscale{1}
\caption{\label{landmovie}
Left columns: Overhead views of mass, energy, and vertical momentum
deposited in Jupiter's atmosphere by an SL9 impact plume, over time.
This model run has no sliding.  Each landing quantity has the same
log-scaled grayscale throughout its column.  Early landing views are
saturated so that later views can be seen at all.  The patterns and
deposition peaks differ.  Mass peaks early, energy is strong
throughout, and momentum peaks in mid-plume.  The panel width is
12,000 km.  Right column: 3D rendering of a plume in flight, viewed
from above the southwest.  The rendering shows a yellow isosurface of
mass density (in this case nearly a material surface) and two planes
showing mass density.  These planes are the \math{z}=0 level and the
vertical plane containing the plume axis.  The latter has been colored
blue at zero density as an aid to viewing.  Rendering only inside the
indicated box makes the interior density gradient visible.
}
\end{figure*}

\begin{figure*}[tbp]
\plotone{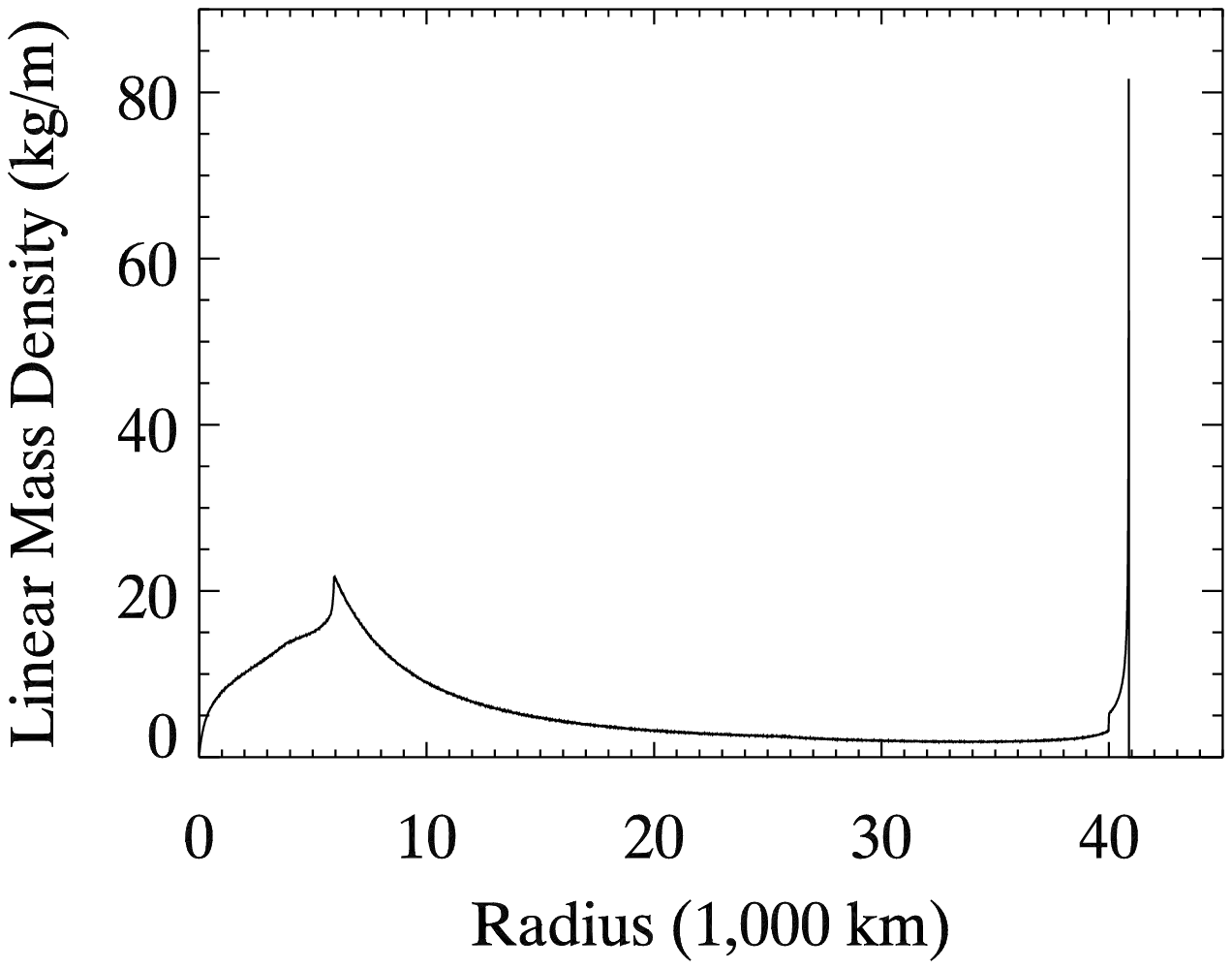}
\caption{\label{radmass}
Mass as a function of radius, integrated over azimuth, after sliding
has stopped.  Definite integrals of this function give the mass in
each part of the debris field.
}
\end{figure*}

\begin{figure*}[tbp]
\plotone{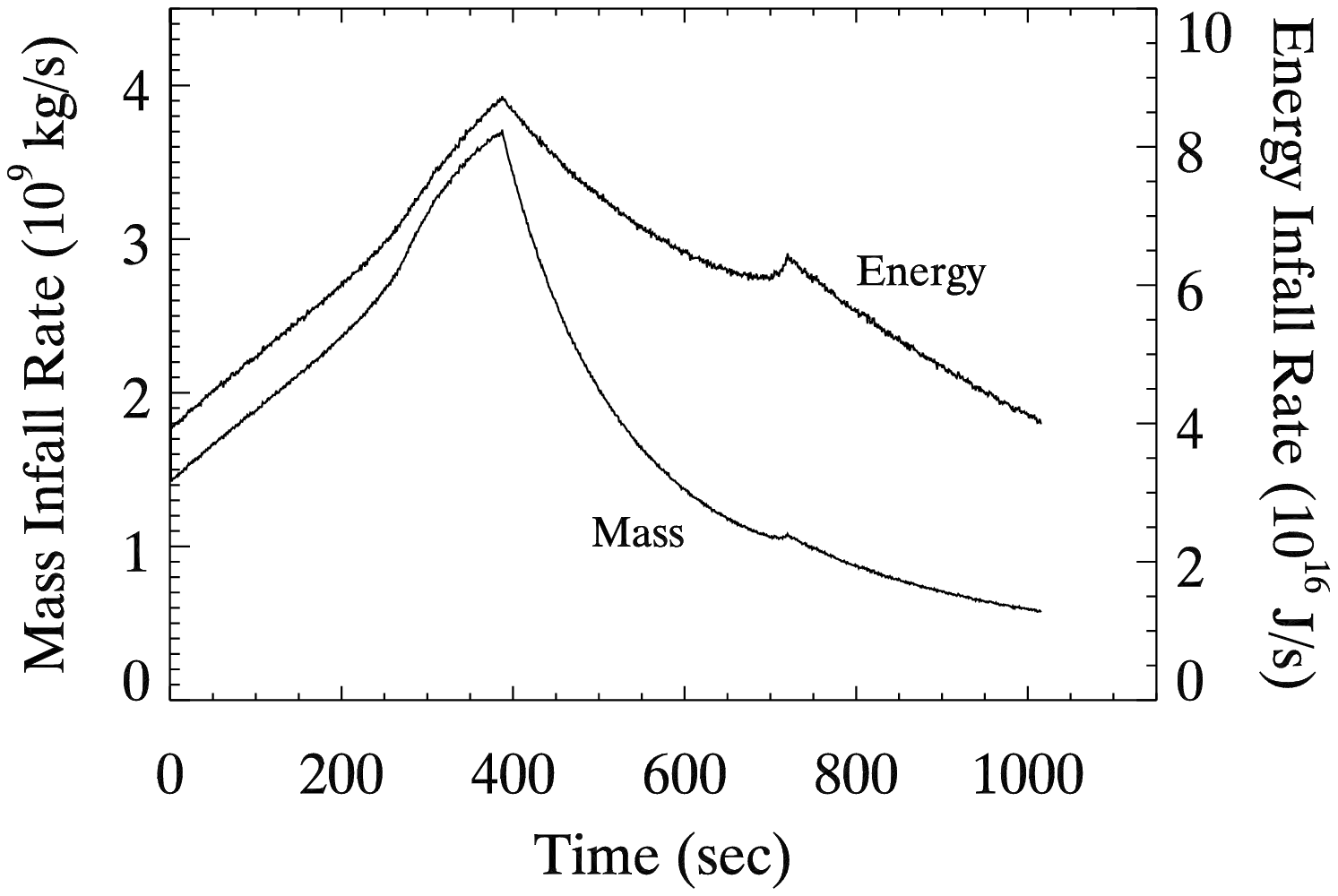}
\caption{\label{timemass}
Rate of total mass and energy impacting the atmosphere (no sliding) as
a function of time.  Compare to the lightcurve shown in Fig.\ 8 of
Paper II.  The spike at 740 sec corresponds to the onset of CO
emission observed by \citet{meaceslw}.
}
\end{figure*}

The distribution of mass, kinetic energy, and momentum flux in the nominal
plume infall is sharply time-dependent.  Fig.\ \ref{landmovie} shows the
three quantities as well as a 3-dimensional view as a function of
time.  The main event, which peaks at 600 sec
\citep{nicghslrlcgrl}, is radiation due to the arrival of kinetic
energy.  The lightcurves' PC3 corresponds to the early,
nearly-horizontal expansion of
the vanguard through the atmosphere.  The final collapse of the
vanguard around 1000 sec causes the transient flare in the
\sim0.9{\micron} lightcurves.  Paper II covers both topics in more
detail.  Figure \ref{radmass} shows the azimuthally-integrated mass
distribution after all material has stopped.  Figure \ref{timemass}
gives the mass and kinetic energy fallen as a function of time.

\comment{\placefigure{landmovie}}

According to Eq.\ \ref{kzcompcummass}, a plume entrains 10--100
impactor masses of Jovian air.  The uncertainty comes from the unknown
\math{\eta} and the question of what low-end velocity to choose as the
cutoff value for the plume (without which it contains infinite mass).
With ZM95's arbitrary \math{\eta} = 0.3 and {\vmin} = 1 km/s, a plume
entrains 80 impactor masses of air.  The amount of entrainment scales
linearly with \math{\eta}.  For {\vmin} = 2 km/s, the entrainment is
27 \math{m\sb{i}} and for {\vmin} = 4.5 km/s it is 7.8 \math{m\sb{i}}.
\citet{ahrtooisljgrl} and \citet{takoaslsijic} calculate values of 20
and 40 \math{m\sb{i}} for their 2-km fragment.  Even with the extremes
of temperature and pressure experienced by plume gas, molecular
hydrogen is so dominant that values for the molecular weight of
infallen plume material should be close to that of Jovian air.

\comment{\placefigure{radmass}}

\comment{\placefigure{timemass}}

\citet{craslmfpfeiau} present evidence in their Fig.\ 9 that the
visible portion of the plumes are due to a lofted cloud deck, and that
there may be invisible portions that went many times higher.  We set
\math{\vmax} based on the observed \sim3,000 km maximum altitudes
\citep{jescbhslplic},
but could set it arbitrarily and adjust sliding parameters
accordingly.  However, in Paper II we present a lightcurve match based
on our nominal parameters.

\section{Conclusions}
\label{concl}

We initialize a plume model with the final conditions of an
entry-response model, preserving the physical state as much as
possible as we transition from one set of dominant physics to the
next.  In so doing, we have discovered physical explanations for a
number of disparate, previously-unexplained phenomena, including
McGregor's ring, PC3, and the flare.  All of these
phenomena depend on the details of the geometry and the initial
conditions: reduced-dimension models and most ``reasonable'' assumed
initial conditions will not reproduce them. 

The parameter phase space for SL9 impact plumes is large.  However, by
using the ballistic approximation, we can determine reasonable values
for the geometry and the velocity distribution.  Comparison of our
best debris-field view to the actual data still leaves us short of
perfection.  The differences are likely due to non-ballistic effects
in the very young plume, the details of grain formation, and the
hydrodynamic details of post-reentry silding.  Models that can
calculate these effects are much more complicated than this one, and
exploring phase space with such a model is computationally
prohibitive.  However, those with such models may safely restrict
their investigations to geometry, velocity distribution, and sliding
that is consistent with the nominal parameters given in Table
\ref{paramtab}.  We encourage workers who wish to use the quantities
presented here to contact us for numerical versions.

We leave a number of questions for future work.  Modifications to take
into account the planet's roundness and rotation would delay infall of
material away from the impact site.  Coriolis effects in flight and
in sliding would rotate the patterns.  Finally, there are likely still
things to learn from direct analysis of the plumes in flight (see
Fig.\ \ref{hstplume}), especially when compared to models.

We postulate a particular velocity distribution to explain the
features of the SL9 impact sites.  Images of the sites are static
(except for the waves), and possibly other distributions could produce
the same or better pictures.  However, we also have timing
observations in the forms of both light curves and spectra, and we
have not tuned any plume parameters to match them, except to set
{\mvmin} as suggested by \citet{carwsstsasleso}.  In Paper II we take
up a thorough test of the plume presented here, dropping it onto a
radiative-hydrodynamic atmosphere model that creates synthetic light
curves.  We compare those light curves to the observations, making
this the first detailed model of plume collapse on an atmosphere that
is constrained by observation of an actual collapsing plume.

Although SL9 was a once-in-a-lifetime event, lifetimes are short, and
impacts are common and important in our solar system's history.  We
view the SL9 impacts not as a one-time curiosity, but as a vital
window on processes that shaped the Earth and planets.  We are excited
by the prospect of the next large plume event, and entertain the
notion that it may be man-made.

\acknowledgments

We thank P.\ J.\ Gierasch, M.-M.\ Mac Low, P.\ D.\ Nicholson, and K.\
Zahnle for numerous discussions.  H.\ Hammel and J.\ Mills provided
image processing information.  A portion of this work was performed
while J.\ H.\ held a National Research Council-NASA Goddard Space
Flight Center (GSFC) Research Associateship.  This work was supported
by NSF grant AST-9526314 and by the NASA Planetary Atmospheres and
Planetary Astronomy programs.  We used software and services from the
NASA Astrophysics Data System, GSFC, the Space Science and Engineering
Center, University of Wisconsin - Madison, and the Free Software
Foundation.

\nojoe{\bibliography{slplume1-20-nobigfig}}

\end{document}